\pdfoutput=1 
\documentclass[
 floatfix,
superscriptaddress,
 showkeys,
 aps,
reprint
]{revtex4-1}
\usepackage{mathtools}
\usepackage{array}
\usepackage{tabularx}
\usepackage{graphicx}
\usepackage{dcolumn}
\usepackage{bm}
\usepackage{amsmath}
\usepackage{color}
\usepackage{wasysym}
\usepackage{float}
\usepackage{array,url,kantlipsum}
\usepackage{verbatim}
\usepackage{xcolor}
\usepackage{CJK}
\usepackage{textcomp}
\newcolumntype{P}[1]{>{\centering\arraybackslash}p{#1}}
\usepackage{array}
\usepackage{amssymb}
\newcolumntype{x}[1]{>{\centering\arraybackslash\hspace{0pt}}p{#1}}

\begin{document}


\title{High homogeneity permanent magnet for diamond magnetometry}

\author{Arne Wickenbrock}
\affiliation{Helmholtz-Institut, GSI Helmholtzzentrum f{\"u}r Schwerionenforschung, 55128 Mainz, Germany}
\affiliation{Johannes Gutenberg-Universit{\"a}t Mainz, 55128 Mainz, Germany}
\author{Huijie Zheng}
\affiliation{Helmholtz-Institut, GSI Helmholtzzentrum f{\"u}r Schwerionenforschung, 55128 Mainz, Germany}
\affiliation{Johannes Gutenberg-Universit{\"a}t Mainz, 55128 Mainz, Germany}
\author{Georgios Chatzidrosos}
\affiliation{Helmholtz-Institut, GSI Helmholtzzentrum f{\"u}r Schwerionenforschung, 55128 Mainz, Germany}
\affiliation{Johannes Gutenberg-Universit{\"a}t Mainz, 55128 Mainz, Germany}
\author{Joseph Shaji Rebeirro}
\affiliation{Helmholtz-Institut, GSI Helmholtzzentrum f{\"u}r Schwerionenforschung, 55128 Mainz, Germany}
\affiliation{Johannes Gutenberg-Universit{\"a}t Mainz, 55128 Mainz, Germany}
\author{Tim Schneemann}
\affiliation{Helmholtz-Institut, GSI Helmholtzzentrum f{\"u}r Schwerionenforschung, 55128 Mainz, Germany}
\affiliation{Johannes Gutenberg-Universit{\"a}t Mainz, 55128 Mainz, Germany}
\author{Peter Blümler}
\affiliation{Johannes Gutenberg-Universit{\"a}t Mainz, 55128 Mainz, Germany}

 \date{\today}


\begin{abstract}

Halbach magnets are a source of homogeneous magnetic field in an enclosed volume while keeping stray fields at a minimum. Here, we present the design, construction, and characterization for a 10\,cm inner diameter double Halbach ring providing a homogeneous ($<$\,100\,ppm over $1.0\times1.0\times\,0.5\, \text{cm}^3$) magnetic field of around 105\,mT, which will be used for a diamond based microwave-free widefield imaging setup. The final characterization is performed with a novel fiberized diamond-based sensor on a 3D translation stage documenting the high homogeneity of the constructed Halbach array and its suitability for the proposed use.
\end{abstract}


\maketitle
\section{Introduction}
Cylindrical Halbach magnets \cite{Halbach1980} have become a standard design for nuclear magnetic resonance (NMR) at field strength typically below 2 tesla \cite{PeterBook}.  Depending on the ratio of inner and outer diameter, they can generate very homogeneous magnetic flux densities. They are particularly attractive because their design has the highest flux to mass ratio \cite{Samofalov2006} and they essentially produce no stray fields. Halbach magnets can also be designed to allow a high degree of optical or mechanical access (e.g. for lasers, optics, mechanical parts or very long samples), which is a requirement for certain experiments.
In conjunction with magnetometry based on nitrogen-vacancy (NV) centers in diamond a background field of 102.4\,mT allows for magnetic sensitivity without the need for microwaves \cite{wickenbrock2016microwave,levelanticrossing}, which are required at most other magnetic background fields. A homogeneous magnetic field of this magnitude over a certain volume allows for microwave-free imaging of magnetic structures on top of an NV center hosting diamond sample.

\begin{figure*}
    \centering
    \includegraphics[width=\textwidth]{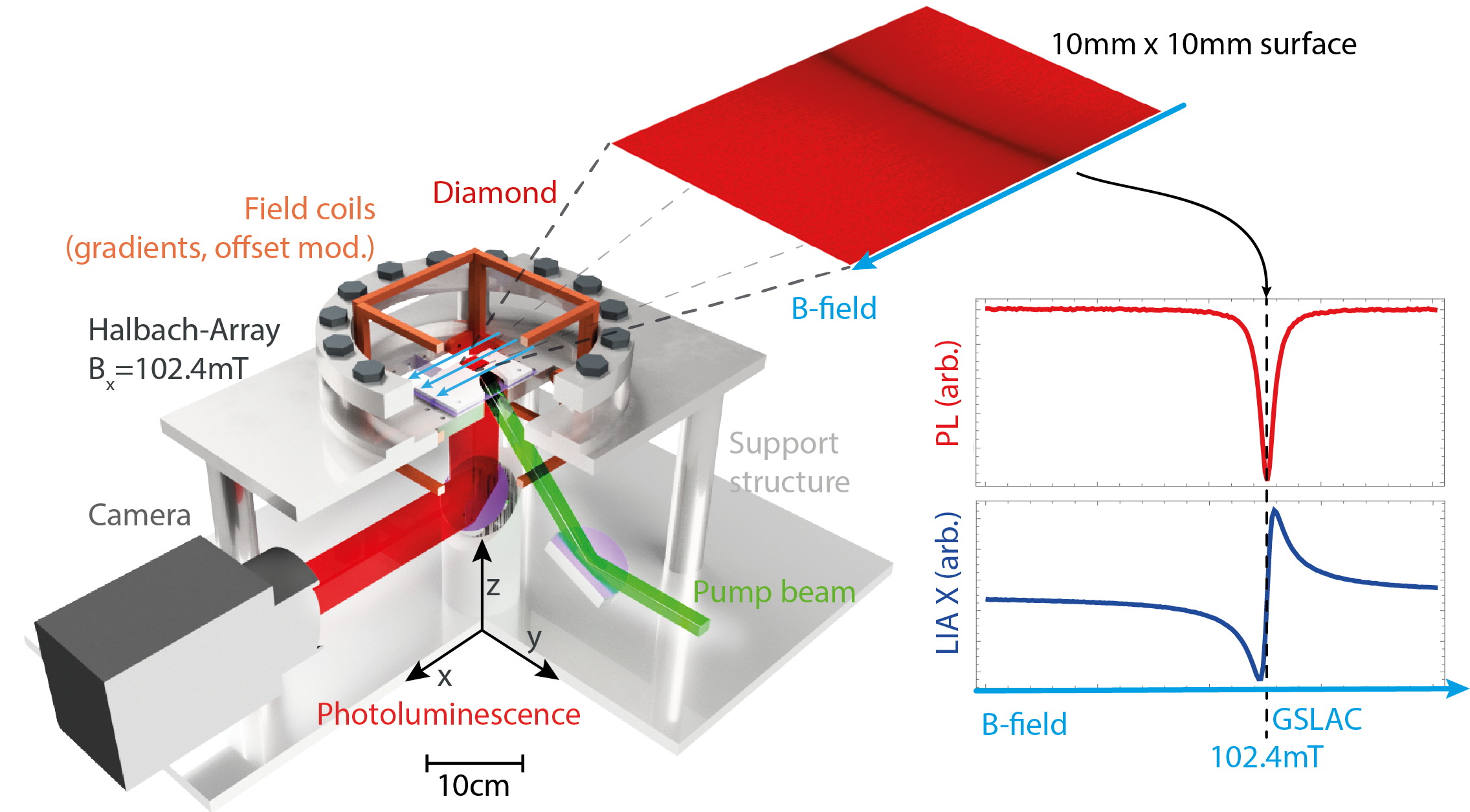}
    \caption{Schematic of a proposed experiment using the described Halbach magnet: a large diamond with a thin layer of nitrogen vacancy centers is placed in a homogeneous 102.4\,mT magnetic field. At this field, the so-called ground-state level anticrossing (GSLAC) occurs,  and the red photoluminescence (PL) of the diamond under illumination with a green 532\,nm pump laser beam depends on the local magnetic field. Small changes in the local magnetic field from a sample placed on top of the diamond can now be directly imaged with an objective and a camera. Providing millions of simultaneous magnetic field measurements with potential diffraction limited resolution. Additional field coils can be used to shim the field (e.g. from the 105\,mT to the working point at 102.4\,mT i.e. the field corresponding to the GSLAC), apply gradients or oscillating fields to probe different aspect of the magnetic properties of the sample and/or allow demodulation with a lock-in amplifier.}
    \label{fig:GSLACSchematic}
\end{figure*}

Recent magnetometry-protocol developments even allow for microwave-free vector-magnetic field measurements using just a single axis NV center ensemble \cite{Zheng2020}. 
Wide-field imaging of magnetic phenomena is an established technique in the field of diamond magnetometry and requires applied oscillating magnetic fields at around 2.8\,GHz. The range of application goes from imaging magnetic bacteria in a biology context \cite{LeSage2013widefield} to probing remanent magnetization of small rock samples to study the history of geophysical magnetic fields \cite{Walsworth2017QDM}. Microwaves can have a detrimental effect on living samples and get severely altered in the presence of conductive materials. However, oscillating fields can also be used to probe conductivity, which is commonly done with coils, but has recently become more popular in a low frequency regime with atomic magnetometers
\cite{Wicken2014eddycurrent} as well as NV center based sensors \cite{Chatzi2019eddycurrent}. With a microwave-free magnetic wide-field imaging setup, conductivity maps can be produced conveniently with high spatial resolution. A schematic of the proposed experiment can be seen in Figure \ref{fig:GSLACSchematic}. A large area diamond containing NV centers is placed in a background field of 102.4\,mT. A camera can then be used to directly read-off magnetic field variation originating from samples placed on the top of the diamond surface. Deploying the protocols developed in \cite{Zheng2020} would allow full 3D vector magnetic field images.
The major component of the magnetic bias field is supposed to be supplied by a Halbach magnet. These magnets provide very stable and highly homogeneous fields and avoid strong electrical currents and associated cooling problems. The fine tuning can then be done by small resistive coil arrangements (cf. Fig. \ref{fig:GSLACSchematic}).

From the perspective of magnet design, the request of a bias-field of defined strength and high homogeneity is somewhat different from the typical NMR-approach, where the field has to be as strong as possible for given constraints. Therefore, we do review the calculation of the resulting magnetic flux in Halbach multipoles and compare the homogeneity of different discrete designs. As suggested in \cite{Peter3}, the best homogeneity can be obtained by avoiding pronounced edges in the magnetic material. For this reason, the homogeneity of a magnet assembled from pieces improves with the order of the polygons used. As the limiting case deploying cylinders, spherocylinders or spheres result in even better performance. The latter have the disadvantage that their magnetization direction can not be aligned with a geometrical feature. Hence, it is very likely that alignment errors are introduced during magnet construction. 
In \cite{Chandrana} this problem is solved by aligning cylindrical magnets in a Halbach configuration by an anchor magnet which represents the line dipole whose stray field defines the orientation of the magnetization in an ideal Halbach \cite{Coey_book}. However, the mutual interaction between the cylindrical magnets sets a principal limit to the achievable homogeneity by this approach. Nevertheless, we used the idea of orienting magnetic cylinders in the field of another (orientation) magnet, and then fixate them with well-defined orientation inside a suitable support structure. These composite elements are then used to assemble the final magnet with high precision.

The manuscript is structured as follows: at first we present in detail the design ideas behind this Halbach magnet and how the flux can be calculated for realistic assemblies. The problem to predict the target field of a given magnet assembly is reviewed and solved in detail because we are not aware of a comprehensive summary in the literature. Furthermore, the analytical treatment allows to create seed designs, that then can be refined in numerical simulations. We then compare different magnet geometries and designs in terms of homogeneity and field strength. In the following section, we document the construction process from the individual permanent magnet pieces to the final assembled double Halbach ring magnet. Finally, the magnet is characterized with a fiberized diamond vector magnetometer moved via a 3D translation stage in the volume of the assembly. The resulting central homogeneity of $7\mu$T (67 ppm) at a central field around 104.5\,mT is confirmed with NMR measurements and suitable for the suggested widefield imaging experiment. 

\section{Design Concept}
\label{sec:concept}
\subsection{Halbach magnets}
In order to avoid confusion we first want to define a nomenclature. The final assembly is called ``Magnet". It consists of several smaller, permanent magnets, referenced as ``PM". If the PMs have an identical shape, magnetization strength and direction the resulting magnet is called ``Mandhala" (Magnetic Arrangement for Novel Discrete Halbach Layout \cite{Peter1}).

In this section we review the equations to calculate the magnetic field of Halbach cylinders and particularly Mandhalas and check their validity by comparison with finite element (FEM)-simulations.

\begin{figure}
    \centering
    \includegraphics[width=0.5\columnwidth]{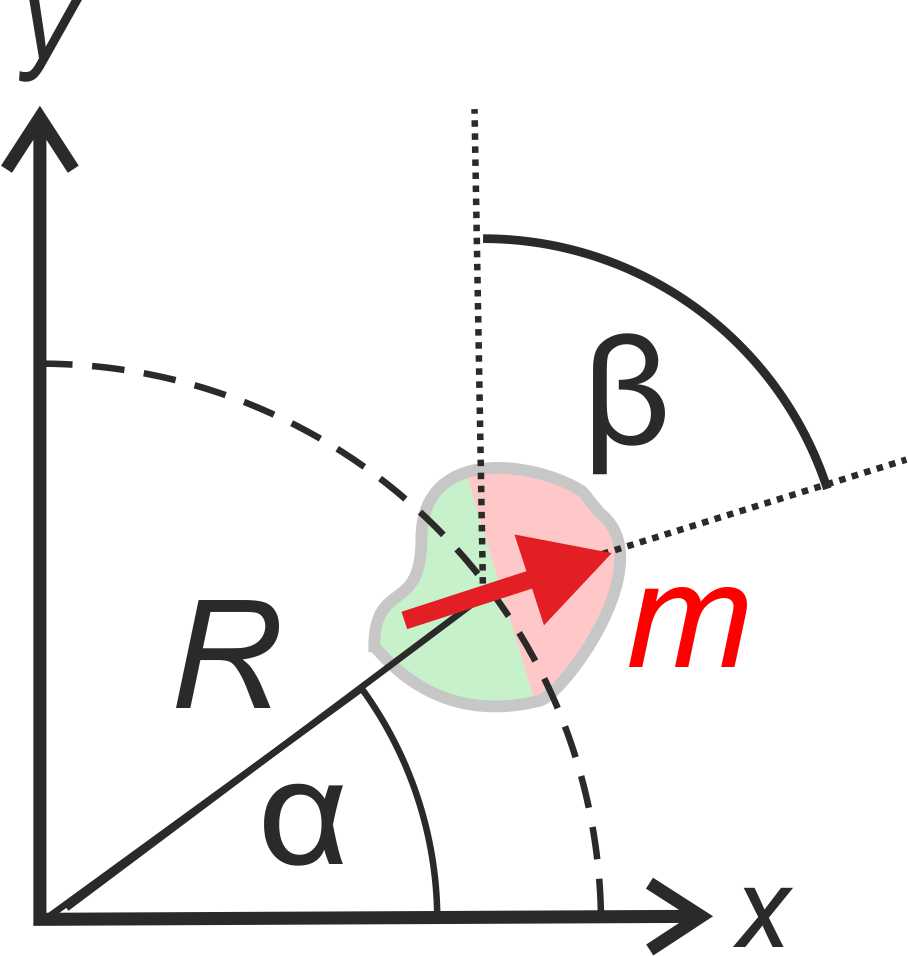}
  \caption{The magnets are positioned on a circle of radius $R$ at an angle $\alpha$ with respect to the $x$-axis. Their magnetization (red arrow) is oriented at an angle $\beta$ which is a multiple of $\alpha$ (cf. eq. \eqref{eq:1}.}    
  \label{fig:0}
\end{figure}

Cylindrical Halbach magnets consist of magnetization which is arranged around a center at a position angle, $\alpha$, while the magnetization points at a magnetization angle (see Fig. \ref{fig:0}) , 
\begin{equation}
    \beta=(1+k) \alpha \text{  with  }k\in  \mathbb{Z} \setminus \{0\}
    \label{eq:1}
\end{equation}
which is an integer multiple of $\alpha$. The constant $k$ determines the polarity $p=2k$ of the resulting magnetic field. The sign of $k$ determines if the entire flux will be inside ($k>0$) or outside ($k<0$) the magnetic structure \cite{Peter1}. The most interesting settings for NMR-magnets are first of all $k=+1$, resulting in an inner dipole ($p=2$) with a homogeneous field inside and no outer stray field. However, for spatially encoding experiments ($\vec{k}$- or $\vec{q}$-space) $k=+2$, resulting in a quadrupolar field ($p=4$), is very interesting since very strong magnetic field gradients can be generated this way \cite{Peter2}. In ideal Halbach magnets \cite{Halbach1980}, the direction of the magnetization $\vec{m}$ rotates continuously in the magnetic material at a fixed radius $R$, 
\begin{equation}
   \tilde{m}\left(\tilde{r}\right)=B_R e^{\tilde{i}\beta}=B_R e^{\tilde{i}\left(1+k\right)\alpha} \textrm{   with   }\tilde{r}=R e^{\tilde{i}\alpha}
   \label{eq:2}
\end{equation}
where the tilde indicates complex variables and $B_R$ is the remanence of the magnetic material. As long as cylinders cannot be magnetized with such properties, real Halbach systems have to be made from discrete PM pieces and many different designs have been discussed to do so \cite{PeterBook}. They consist either from cylindrical or wedge-shaped segments with individual magnetization directions or from identical pieces which are mounted in a supporting structure at individual angles, the latter is also known as a Mandhala \cite{Peter1}. The main advantage of Mandhalas is that the PMs are identical. This allows for ordering them by magnetic remanence $B_R$, since random variation in $B_R$ would otherwise reduce the homogeneity of the composite magnet. Theoretically the flux inside ideal Halbach cylinders with $k>1$, an inner radius $R_i$ and an outer radius $R_o$ is given by \cite{Halbach1980}: 

\begin{align}
    \tilde{B}^*\left(\tilde{r}\right)&=B_x\left(\tilde{r}\right)-\tilde{i} B_y\left(\tilde{r}\right)=f \tilde{r}^{k-1}  \label{eq:3}\\
    \text{with  } f&=B_R\ln{\frac{R_o}{R_i}} \quad\textrm{for  }k=1 \nonumber\\
    \textrm{and  } f&=B_R\frac{k}{k-1}\left(\frac{1}{R^{k-1}}-\frac{1}{R_o^{k-1}}\right)\quad\textrm{for  }k>1\nonumber
\end{align}
where $\tilde{r}=x+\tilde{i} y$ is a complex defined space variable ($\left|r\right|<R_i$) and the asterisk indicates the complex conjugate\footnote{Here we follow the original description by K.\,Halbach \cite{Halbach1980}} (coordinate system: the main flux component defines the $x$-direction, $y$ is orthogonal to $x$ and in the cylinder plane and $z$ is along the cylinder axis).

The flux in discrete Halbach structures made from identical PMs can be calculated analytically. 
This generates intuitions about the geometrical dependencies and provides starting values for finite element simulations. We consider three different options for the PM's cross-section assuming that the hollow Halbach cylinder is assembled from $N$ identical magnets (having the same shape and direction of magnetization, i.e. a Mandhala \cite{Peter1}): a) circular, b) squared or c) polygonal with $N/2$ vertices. Table \ref{table:1} summarizes the resulting geometrical constraints and introduces the inter-magnet distance, $d$, which is zero for a dense packed Mandhala but might be chosen $d>0$ e.g. for allowing support structures between neighboring magnets \cite{Peter3}.

\begin{table*}[ht]
\centering
\begin{tabular}{|p{0.15\textwidth}|>{\centering}p{0.27\textwidth}|>{\centering}p{0.27\textwidth}|>{\centering\arraybackslash}p{0.27\textwidth}|}
    \hline
    Property & Circles & Squares & $N/2$-Polygon  \\
    \hline
    \hline
    magnet size $a$ &  $2R \sin\frac{\pi}{N}-d$ & $2R\Xi\left(N\right)-d$ & $2R\Theta\left(N\right)-d$  \\
    inner Radius $R_i$ &  $R \left(1-\sin\frac{\pi}{N}\right)=R-\frac{a+d}{2}$ & $R\left(1-\sqrt{2}\Xi\left(N\right)\right)=R-\frac{a+d}{\sqrt{2}}$ & $R\left(1-\frac{\Theta\left(N\right)}{\sin{\frac{2\pi}{N}}}\right) = R-\frac{a+d}{2\sin\frac{2\pi}{N}}$\\ 
    outer Radius $R_o$ &  $R \left(1+\sin\frac{\pi}{N}\right)=R+\frac{a+d}{2}$ & $R\left(1+\sqrt{2}\Xi\left(N\right)\right)=R+\frac{a+d}{\sqrt{2}}$ & $R\left(1+\frac{\Theta\left(N\right)}{\sin{\frac{2\pi}{N}}}\right) = R+\frac{a+d}{2\sin\frac{2\pi}{N}}$\\ 
    Magnets area $A_M$ &  $N\pi\frac{a^2}{4}$ & $N a^2$ & $\frac{N^2a^2}{16}\cot{\frac{\pi}{N}}$\\ 
    \hline
 \end{tabular}
    \caption{Summary of geometric properties for Mandhalas built from magnets with circular, square, and polygonal footprint. Given the magnet number, $N$, the radius of their centers, $R$, their shape-specific size, $a$, the inner and outer radius of the Mandhala can be calculated for an arrangement where all neighboring magnets have a distance, $d$, to each other. The shape-specific size $a$ corresponds to the diameter for circles, the side length for squares and  $N/2$-polygons. Furthermore the formula to calculate the area of all $N$ magnets, $A_M$, is given (see \cite{Peter4} for details). For a compact representation two geometric abbreviations, $\Xi$ and $\Theta$ are introduced in eqs. \ref{eq:xi} and \ref{eq:theta}.}
    \label{table:1}
\end{table*}

The following abbreviations are used in table \ref{table:1}:
\begin{equation}
    \Xi\left(N\right)\coloneqq\frac{\cos{\left(\frac{2\pi}{N}\right)}-\sin{\left(\frac{2\pi}{N}\right)}-\sqrt{2}\sin{\left(\frac{\pi}{4}-\frac{4\pi}{N}\right)}}{2\cos{\left(\frac{\pi}{4}-\frac{4\pi}{N}\right)}+\sqrt{2}}
    \label{eq:xi}
\end{equation}
and
\begin{equation}
    \Theta\left(N\right)\coloneqq \begin{cases}
               \frac{\sin{\left(\frac{2\pi}{N}\right)}^2}{2\cos{\left(\frac{2\pi}{N}\right)}}       & \text{for }\frac{N}{2}\text{ even}\\
               2\frac{\sin{\left(\frac{\pi}{N}\right)}\sin{\left(\frac{2\pi}{N}\right)}}{2\cos{\left(\frac{2\pi}{N}\right)}+1}       & \text{for }\frac{N}{2}\text{ odd}\\
           \end{cases}
\label{eq:theta}
\end{equation}
The flux of a discrete approximation of an ideal Halbach is reduced by the ratio of its area $A_M$ with respect to the area of an ideal hollow Halbach cylinder (i.e. ${A_M}^{\text{ideal}}=\pi\left({R_o}^2-{R_i}^2\right)$), because the flux of PMs scales with the volume \cite{Camacho}. Hence, eq. \eqref{eq:3} for the flux inside an infinitely long Mandhala has to be modified:
\begin{equation}
    f' = \frac{A_M}{{A_M}^{\text{ideal}}}f = \frac{A_M}{\pi\left({R_o}^2-{R_i}^2\right)}f
    \label{eq:4}
\end{equation}
The validity of equation \eqref{eq:4} is confirmed by 2D-FEM simulations (COMSOL Multiphysics 5.5) and summarized in table \ref{table:2}. The analytically calculated values are in good agreement (4-11 \%) with the FEM-simulations.

\begin{table}[ht]
\centering
\begin{tabular}{|p{0.08\textwidth}|>{\centering}p{0.08\textwidth}|>{\centering}p{0.07\textwidth}|>{\centering}p{0.07\textwidth}|>{\centering}p{0.07\textwidth}|>{\centering\arraybackslash}p{0.07\textwidth}|}
    \hline
    shape & $a\,[\text{mm}]$ & $B^a_0\,[\text{mT}]$ & $B^s_0\,[\text{mT}]$ & $\Delta\,[\text{ppm}]$ & $\sigma\,[\text{ppm}]$  \\
    \hline
    \hline
    segmented & 32 & 533 & 507 & 0.44 & 0.08  \\
    \hline
    squares & 26 & 338 & 325 &103 & 21   \\
    \hline
    octagons & 13.25 & 441 & 408 & 0.31 & 0.06  \\
    \hline
    cylinders &  32 & 402 & 356 & 0.32 & 0.06 \\
    \hline
 \end{tabular}
    \caption{Comparison of the analytically calculated field strength, $B^a_0$ from eq.\,\eqref{eq:4} with the results of a 2D-FEM simulation performed in COMSOL 5.5 for infinitely long Halbach cylinders discretized in $N=16$ segments of different shape: the size, $a$, for squares means the side length, for the regular octagons $a$ is a single side [the span is $(1+\sqrt{2})a$] and the diameter for cylinders all arranged with their centers on a circle of radius $R=85$\,mm and a remanence $B_R = 1.4$\,T. $B^s_0$ is the simulated magnetic flux in the center and $\Delta$ is the difference of maximum and minimum $B^s_0$ value, while $\sigma$ is its standard deviation both determined over a central circle with a diameter of 20\,mm and expressed as ppm of $B_0$. The segmented but otherwise ideal Halbach in the first row has $R_o= 101$\,mm, $R_i = 69$\,mm.}
    \label{table:2}
\end{table}

The values in table \ref{table:2} are for infinitely long cylinders and hence unrealistic. If they are truncated to a height $L$, the central field in this cylinder can be estimated as shown in eq. \eqref{eq:5}, effectively by yet another scaling factor for eq. \eqref{eq:3} \cite{Peter4}: 
\begin{equation}
    f'' = \frac{L\left(6R^2+L^2\right)}{\left(4R^2+L^2\right)^{3/2}}f'.
    \label{eq:5}
\end{equation}
However, not only the flux decreases by truncation but also the homogeneity suffers. As shown in \cite{Peter3} the homogeneity can be improved if at least two truncated cylinders are coaxially aligned with their centers at a distance of $0.412\,R$. Additionally, this increases the flux by a factor of 1.36 compared to the value from a single cylinder as calculated by eq. \eqref{eq:5}. Table \ref{table:3} shows the same quantities as in table \ref{table:2} but for systems of two Mandhalas, each truncated to a length of 32\,mm and separated by a gap $g$. The value for $g$ was optimized for highest homogeneity. For the analytical results $g=(0.412\,R–L)=38.04$\,mm is used. The analytical $B^a_0$-values from eqs. \eqref{eq:3} -\eqref{eq:5} agree very well ($<$ 5\%) with the results, $B^s_0$, from a 3D-FEM simulation as shown in table \ref{table:3}.

\begin{table*}[ht]
\centering
\begin{tabular}{|p{0.13\textwidth}|>{\centering}p{0.11\textwidth}|>{\centering}p{0.11\textwidth}|>{\centering}p{0.11\textwidth}|>{\centering}p{0.11\textwidth}|>{\centering}p{0.11\textwidth}|>{\centering}p{0.11\textwidth}|>{\centering\arraybackslash}p{0.11\textwidth}|}
    \hline
    shape & $a\,[\text{mm}]$ &$L\,[\text{mm}]$&$g\,[\text{mm}]$&$B^a_0\,[\text{mT}]$&$B^s_0\,[\text{mT}]$ & $\Delta\,[\text{ppm}]$& $\sigma\,[\text{ppm}]$  \\
    \hline
    \hline
    squares   & 26 & 32 & 38.7 & 126 & 124 & 346 & 49 \\
    \hline
    octagons  & 32 & 32 & 37.9 & 165 & 157 & 376 & 50 \\
    \hline
    cylinders & 32 & 32 & 37.8 & 150 & 149 & 199 & 27 \\
    \hline
    spheres   & 32 & 32 & 36.9 & -   & 100 & 214 & 25\\
    \hline
 \end{tabular}
\caption{Comparison of the analytically calculated field strength, $B^a_0$ from eq. \eqref{eq:5} with the results of a 3D-FEM simulation with COMSOL 5.5 for two Mandhala cylinders of height $L$ mounted with a gap $g$ between them. The other geometric properties and extracted values are the same as in table \ref{table:2}. The gap distance $g$ was optimized for best homogeneity in the required volume. The numerical optimization was limited to a spatial accuracy of 100\,$\mu$m}
    \label{table:3}
\end{table*}

As shown, it is relatively easy to predict the flux strength of Halbach/Mandhala cylinders from simple equations. Unfortunately, this is not so easy concerning their homogeneity. However, intuitively, structures with edges result in less homogeneous fields than smooth pieces. Inspecting table \ref{table:2} clearly shows this behavior for squared PM shapes while the results of the octagonal shapes are more comparable to the cylinders. This impression changes when looking at the more realistic values of the 3D-FEMs in table \ref{table:3}. From this we conclude that magnets made from cylindrical or spherical PMs are roughly twice as homogeneous compared to those made from polygonal PMs.

The main reason why most Halbach/Mandhala magnets use PMs with polygonal cross-sections is to avoid angular positioning errors \cite{PeterBook}. 
Orientation errors occurring in the production/magnetization process do not alter the homogeneity of the final magnet assuming this error is the same for all magnets. In this case, the only source of variability stems from the variation in the individual magnet's remanence $B_R$. We therefore measured the individual magnets and included these properties in the design, ultimately resulting in an improved homogeneity \cite{Peter3, Peter4}.

\subsection{Making Mandhalas from cylindrical magnets}

\begin{figure*}
    \centering
     \includegraphics[width=\textwidth]{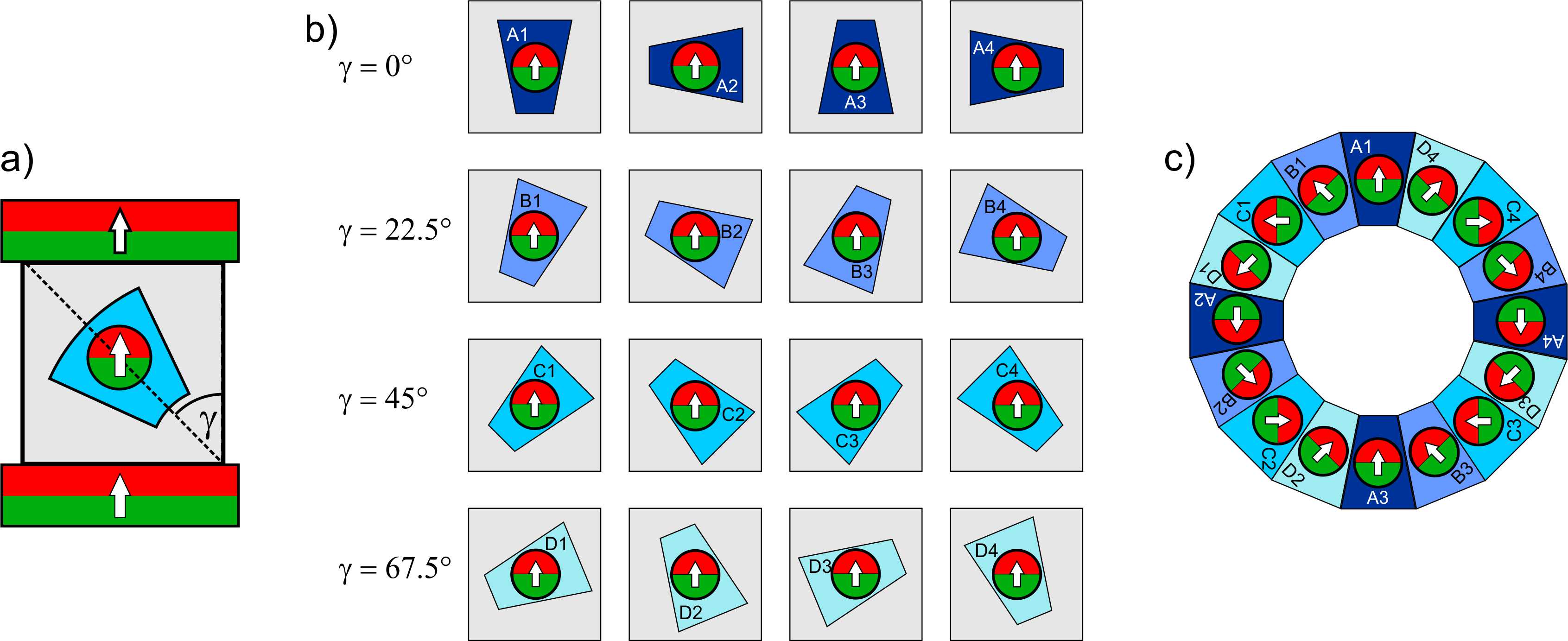}
    \caption{Orientation and assembly of the PMs. a) A round PM is placed in a suitable segment (blue) at an angle $\gamma$ relative to an orientation magnetic field provided by two other (orientation) magnets (top/bottom). The orientation is fixed in space by a mechanical support (gray). b) Illustration how a ($k = 1, N = 16$) Halbach array can be made from just four squared supports each with different $\gamma_j$. In each row the squared support (gray) is rotated by 90$^\circ$ from column to column relative to the external orientation field (indicated by the magnetization arrow in the central PM). Each support hosts a cylindrical segment (different shades of blue per row) at a certain angle $\gamma_j =22.5^\circ (j-1) $ with $j = 1..4$ with a cylindrical PM in its center. The individual segments are annotated by letters A-D for each row and numbers 1-4 for each column. c) Constructed Halbach dipole from the 16 PMs.}
    \label{fig:PMOrientation}
\end{figure*}

From a point of homogeneity it would be advantageous to build Mandhalas from round PMs. Cylinders offer more flexibility over spheres because their height can be varied according to eq. \eqref{eq:5} to adjust the flux strength, but in principle the following concept can also be used for spherical PMs (or combine both in spherocylinders which we didn't consider here). 
The new aspect of this work is that the magnets are magneto-mechanically oriented inside a supporting structure by a constant external magnetic field. Therefore, a cylindrical segment with a hole for the PM is placed at a well-adjusted angle, $\gamma$, relative to an orientation field provided by two other larger and stronger PMs. These are only used for this orientation procedure (see Fig. \ref{fig:PMOrientation}a).

The PM is then fixed in the oriented position by glue. Of course an individual support for each orientation angle could be made, but depending on the number $N$ of PMs in the discrete Halbach and constant $k$ there are symmetries which reduce the total number of necessary orientation angles. So for $N=16$ and $k=1$ a hexadecagonal support could be used and rotated in the orientation field to create all required angles $\gamma$. Although this would reduce the number of required support structures to one, we decided instead to use four squares with different $\gamma$ for simplicity and handling. The four different supports including the cylinder segments at the respective angle $\gamma$ are displayed in Fig.\ref{fig:PMOrientation}b. Each support can be used for four different $\gamma$ via rotation in the orientation field. The assembled magnet can be seen in Fig. \ref{fig:PMOrientation}c (see caption for more detail). 
The orientation angle $\gamma$ for each magnet in the assembly can be calculated as the difference of magnetization angle $\beta$ and position angle $\alpha$ (cf. Figs. \ref{fig:0} and \ref{fig:PMOrientation}a). 
\begin{align}
   \gamma_j &=\beta_j-\alpha_j  \hspace{0.8cm}\\ 
   \text{with }\alpha_j&=j\frac{2\pi}{N} \text{ and }j=1...N\nonumber
   \label{eq:6}
\end{align}
and with eq. \eqref{eq:1}
\begin{equation}
    \gamma_j=(1+k)\alpha_j-\alpha_j=k \alpha_j
    \label{eq:7}
\end{equation}

\section{Construction of the magnet}
This section details the construction process of the magnet: first, we motivate the dimensions of the final design, characterize the indivdual PMs and then describe how to put everything together. As stated in the introduction, the magnet system should generate a magnetic flux density of $(102.4\pm 0.1)$\,mT over a central volume of $1\times 1\times 0.5\, \text{cm}^3=0.5\,\text{ml}$. From the discussed design considerations a system consisting of two layers of 16 cylindrical PMs was chosen. In the following we describe the parameter optimization procedure. The length of the PMs was calculated using eqs. \eqref{eq:3} -\eqref{eq:5} resulting in $L = 21.53$\,mm for the required field strength. Starting from this set of parameters, 3D-FEM simulations were used to optimize the homogeneity and the flux density by varying the width of the gap between the two layers, their diameters as well as the diameter and the length of the individual PMs (both to an accuracy of 100\,$\mu$m). Considering manufacturing constraints, the simulations were restricted to whole millimeter numbers for the dimensions of the PMs.
This resulted in slightly different values: $R = 85.0$\,mm, $a = 26$\,mm, $L = 34$\,mm, $g = 37.8$\,mm for magnets with $B_R = 1.4$\,T  generating a flux of 103\,mT with a homogeneity of 9\,ppm in the anticipated volume.
To construct the magnet 70 PMs (cylinders with  26\,mm diameter, 34\,mm height, made from Nd$_2$Fe$_{14}$B grade N52, with diametrical magnetization direction and a Ni-Cu-Ni coating) were custom made from Spacemagnets Europe GmbH. 

\subsection{Measuring the individual magnets}
\begin{figure*}
    \centering
    \includegraphics[width=1\textwidth]{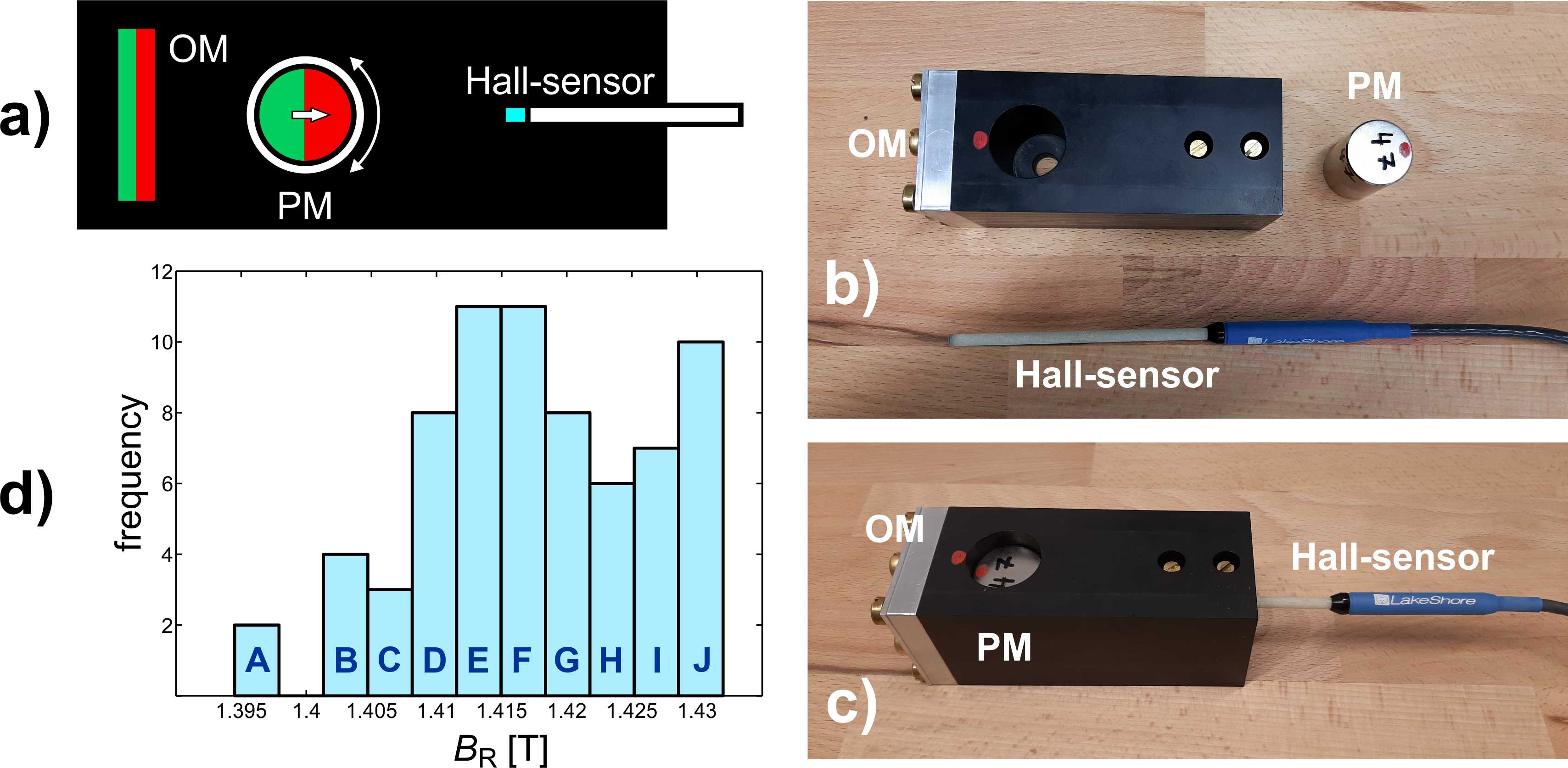}
    \caption{Measuring the strength of individual cylindrical PMs using an orientation device. a) Schematic drawing of the device: The orientation magnet (OM) is a fixed flat permanent magnet to orient the PM via mutual attraction. The PM is inserted in a hole in which it can rotate mechanically. An axial Hall-probe is inserted from the other side to measure the flux of the OM alone or OM+PM combined. b) Photograph of the device made from PVC and aluminium with PM and Hall-sensor dismounted, and c) all components assembled. d) Histogram of the $B_R$-values (each bin labeled with letters from A-J) of the 70\,PMs that were measured. The $B_R$ values were reconstructed from the measured flux values using a 3D-FEM model.}
    \label{fig:PMStrength}
\end{figure*}

Since rare-earth magnets deviate in their properties, it is an established concept in magnet building to measure the strength of the far field of all PMs before construction. Either to select a set with small deviations from the mean or to use their individual remanences to optimize the homogeneity \cite{Peter3}. However, it is difficult to measure the strength of a cylindrical magnet with diametrical magnetization in a single measurement, because the magnetization axis is not coinciding with a geometric feature. Therefore, the magnets must be oriented prior to the measurement. This is done, similar to the concept presented in the design section, by using a another stationary magnet to orient the PM and derive the remanence of each PM from the measurement of their combined flux. We chose a flat block magnet ($30\times30\times5\,\text{mm}^3$, Nd$_2$Fe$_{14}$B grade N45, $B_R = 1.35$\,T, magnetized parallel to the 5\,mm side, EarthMag GmbH) for this purpose. It is located 35\,mm away from the center of a hole in which the cylindrical PMs can be inserted.  To measure the flux, an axial Hall-sensor (HMNA-1904-VF, Lake Shore Cryotronics) is placed 45\,mm away from the center of this hole. 
Everything is centrally aligned. The arrangement can be seen in Figs. \ref{fig:PMStrength}a-c). Without an inserted PM the Hall probe reads a background flux of 1.75\,mT from the orientation magnet only. With inserted PMs the values ranged from 41 to 42\,mT. All 70 PMs were measured this way and their individual remanences inferred from a 3D-FEM simulation of this geometry. The distribution of their remanences, limited by the measurement resolution of the Hall-probe, is shown in Fig. \ref{fig:PMStrength}d with an average of $B_R = 1.416\,\text{T} \pm 0.009 $\,T.

A first guess for an ideal arrangement of the individual PM remanences was calculated from the dipole model in \cite{Peter3}. A more realistic 3D-FEM simulation informed by the measured  remanences led to the final PM-arrangement. The final arrangement is as follows: for the first ring: GGEEGGEEIIGGIIG and the other FJGJGJGJGBFBFBECFB (starting at the A1 position in Fig. \ref{fig:PMOrientation}c and going clockwise with PMs with the following remanences: $\text{B}=1.401\,\text{T},\text{C}=1.405\,\text{T}, \text{D}=\,1.408\,\text{T}$, $\text{E}=1.412\,\text{T}$, $\text{F}=1.415\,\text{T}$, $\text{G}=1.418\,\text{T}$, $\text{I}=\,1.425\,\text{T}$, and $\text{J}\,= 1.429\,\text{T}$ as in Fig.\,\ref{fig:PMStrength}d). This combination of remanences was the optimum for the given PMs. A 3D-FEM simulation of the final arrangement resulted in an average flux of 103.8\,mT and a homogeneity of 27\,ppm in the anticipated volume.

\subsection{Construction of two Halbach rings}
\begin{figure*}
   \centering
   \includegraphics[width=0.8\textwidth]{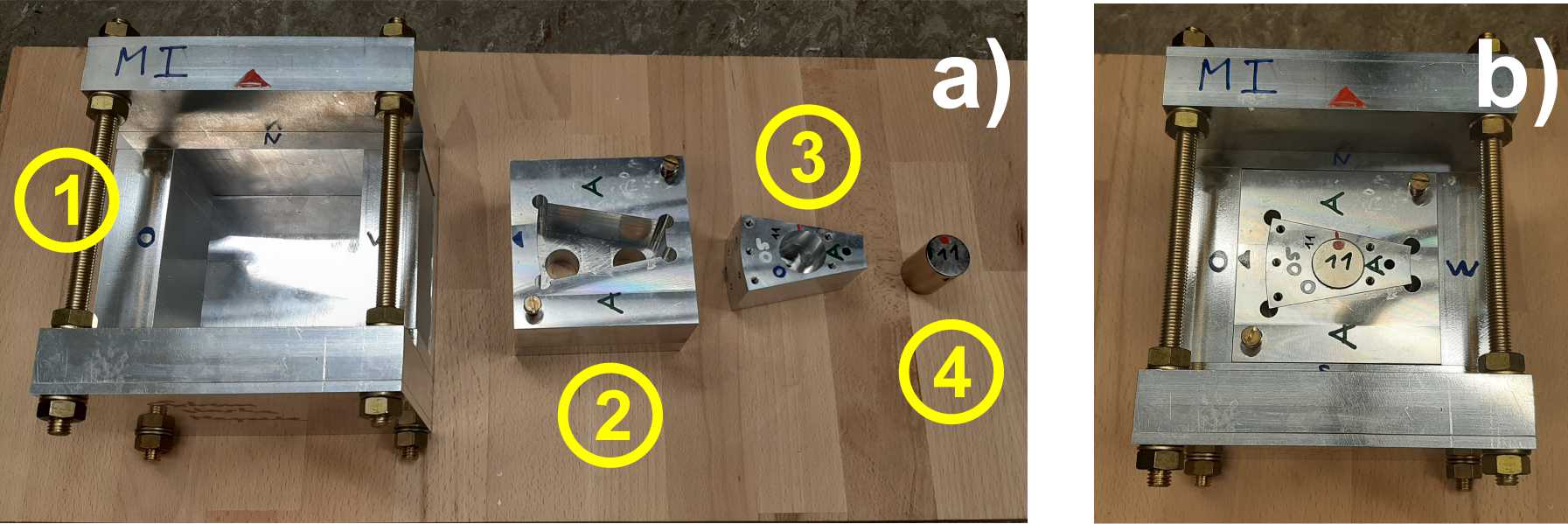}
   \caption{Device to orient PMs in the cylinder segments during fixation: a) all components of the device disassembled: An aluminium box (1) containing the big permanent magnets  ($80\times80\times20\,\text{mm}^3$) for orienting  the PMs. They are mounted and secured in two aluminium boxes which are fixed by threaded 6\,mm diameter brass rods and nuts. In between them is an aluminium block to hold them at a distance of 130 mm. In this block is a socket for insertion of a squared support (2) at four different position (rotation by $90^\circ$). Four of these squared supports exist each for a different orientation angle $\gamma$ (cf. Fig. \ref{fig:PMOrientation}b). Here only the one for $\gamma=0^\circ$ is shown. The squared supports have another socket for insertion of a cylinder segment (3), which has yet another hole in the center into which the PM (4) is  glued. b) same view  with all pieces assembled.}
\label{fig:magnetAssembly}
\end{figure*}
\begin{figure*}
    \centering
    \includegraphics[width=0.8\textwidth]{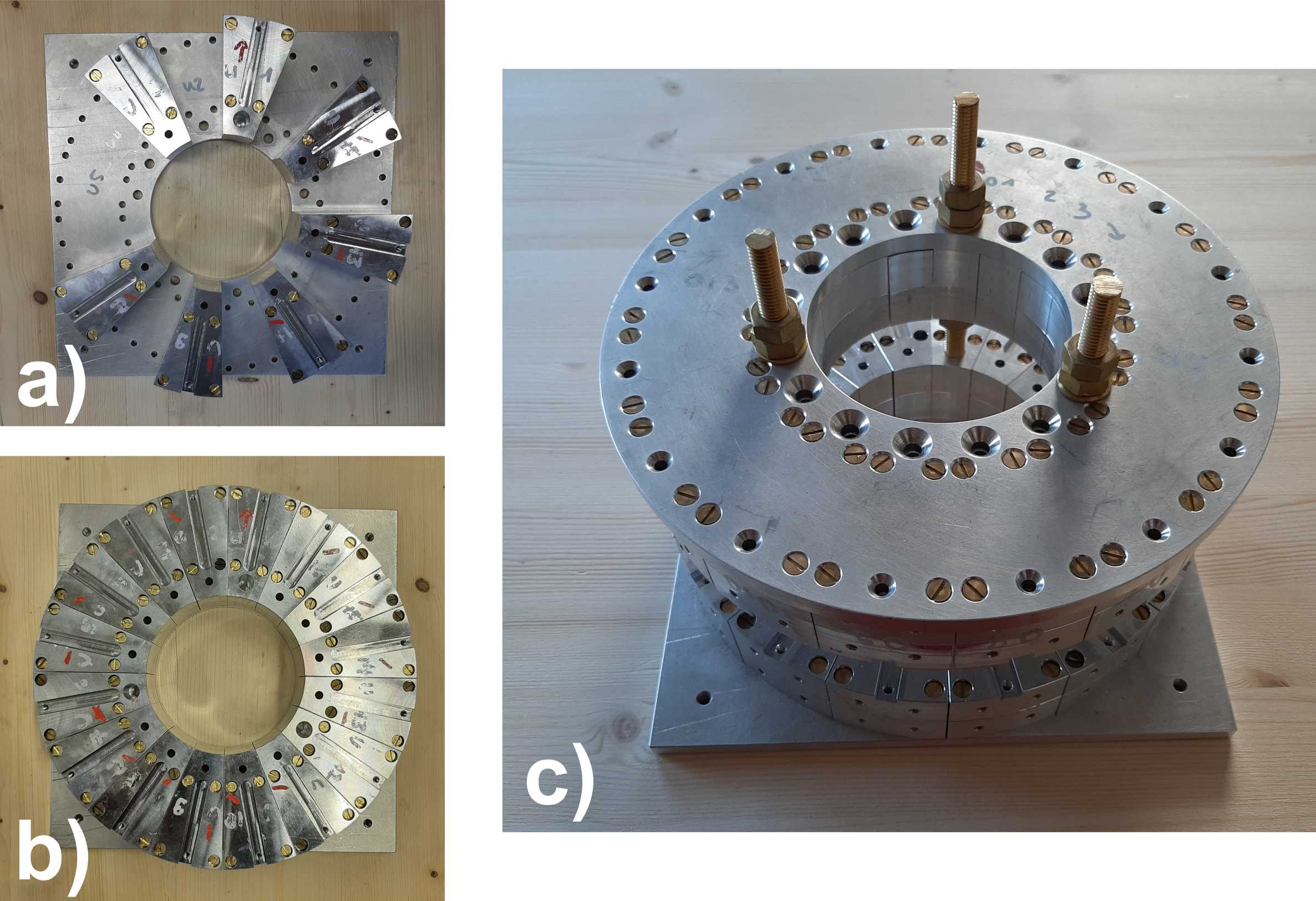}
    \caption{Assembling the magnet: a) Seven cylinder segments being screwed on an aluminium plate. b) Lower half of the magnet finished. c) The completed magnet.}
    \label{fig:5}
\end{figure*}

The construction principle is explained in Fig. \ref{fig:PMOrientation}b/c. Figure \ref{fig:magnetAssembly} shows the mechanical devices used for orienting and glueing the PMs into individual cylinder segments. To orient the PMs, two ($80\times80\times20\,\text{mm}^3$ Nd$_2$Fe$_{14}$B grade N52, $B_R = 1.46$\,T, magnetized parallel to the 20\,mm side, from EarthMag GmbH) are mounted and secured in a special holder such that their magnetization points in the same direction (see Figs.\ref{fig:PMOrientation}a and \ref{fig:magnetAssembly}a). They are separated by a aluminium block with a squared socket for insertion of four different supports in Fig. \ref{fig:PMOrientation}b. Each of these squared supports had another socket in its center to contain a cylinder segment at angles $\gamma = 0^\circ, 22.5^\circ, 45^\circ$, and $67.5^\circ$ (as shown in Fig. \ref{fig:PMOrientation}b). Finally, the cylinder segments have a central hole, into which the individual PMs are glued while being oriented. The construction ensures that the center of the PM coincides with the center of the orientation magnets (further geometrical details are provided in the caption of Fig. \ref{fig:magnetAssembly}).
We used a two-component epoxy resin (UHU Endfest, UHU GmbH) for glueing. In order to lower the viscosity and hardening time the assembled orientation device (Fig. \ref{fig:magnetAssembly}b) was placed into a lab oven at 100$^\circ$C$\pm 5^\circ$\,C for at least one hour. The hardening time of the glue reduces to ca. 10\,min and the bonding strength increases to 25\,N/mm$^2$ at this temperature (compared to 12\,h and 12\,N/mm$^2$ at room temperature). The lower viscosity at higher temperatures is also beneficial for a quick and reliable orientation of the PMs.
In this way, 32 cylinder segments were equipped with PMs and then screwed onto two custom-made aluminium plates in the predefined positions (see Fig. \ref{fig:5}a/b). These two magnet rings were mounted with three 10\,mm threaded brass rods on top each other to form the Halbach magnet (cf. Fig.\,\ref{fig:5}c).The final magnet has an accessible inner hole with a diameter of 100 mm, an outer diameter of 240\,mm, and a height of 122 mm. It weights ca. 9\,kg. and the stray-field at its sides is not exceeding 10\,mT at the surface.


\section{Characterization of the magnet}
The magnetic properties of the assembled Halbach were characterized using three methods. First a coarse measurement of the magnetic field was performed using a Hall-probe sensor on a translation stage, then a nuclear magnetic resonance measurement allowed a more precise estimation of the field homogeneity and magnitude. At last, a highly resolved vector magnetic field map was created using a diamond sensor.

\subsection{Using a Hall probe and NMR}\label{dintro}
The Hall-probe sensor (HMNT-4E04-VR, Lake Shore) was mounted on a computer controlled 3D linear table (OWIS GmbH) and the sample volume, the central $xyz = 10\times10\times5\,\text {mm}^3$, was scanned. 
The magnetic field strength and homogeneity depends on the gap between the two Halbach rings. We chose to fine-tune the gap until a central magnetic field value of around 105\,mT was achieved. Coarse magnetic field measurement scans showed that a slight angle between the two rings improved the homogeneity.
The resulting gap ranges from $g$ = 32.5\,mm to $g$ = 35.5\,mm with the extrema being on the $y$-axis ($x \approx 0$). The best homogeneity is found at 1\,mm below the center. However, the sensitivity of the Hall-probe (0.1\,mT) was insufficient to characterize or homogenize the magnet further.

\begin{figure}
    \centering
    \includegraphics[width=\columnwidth]{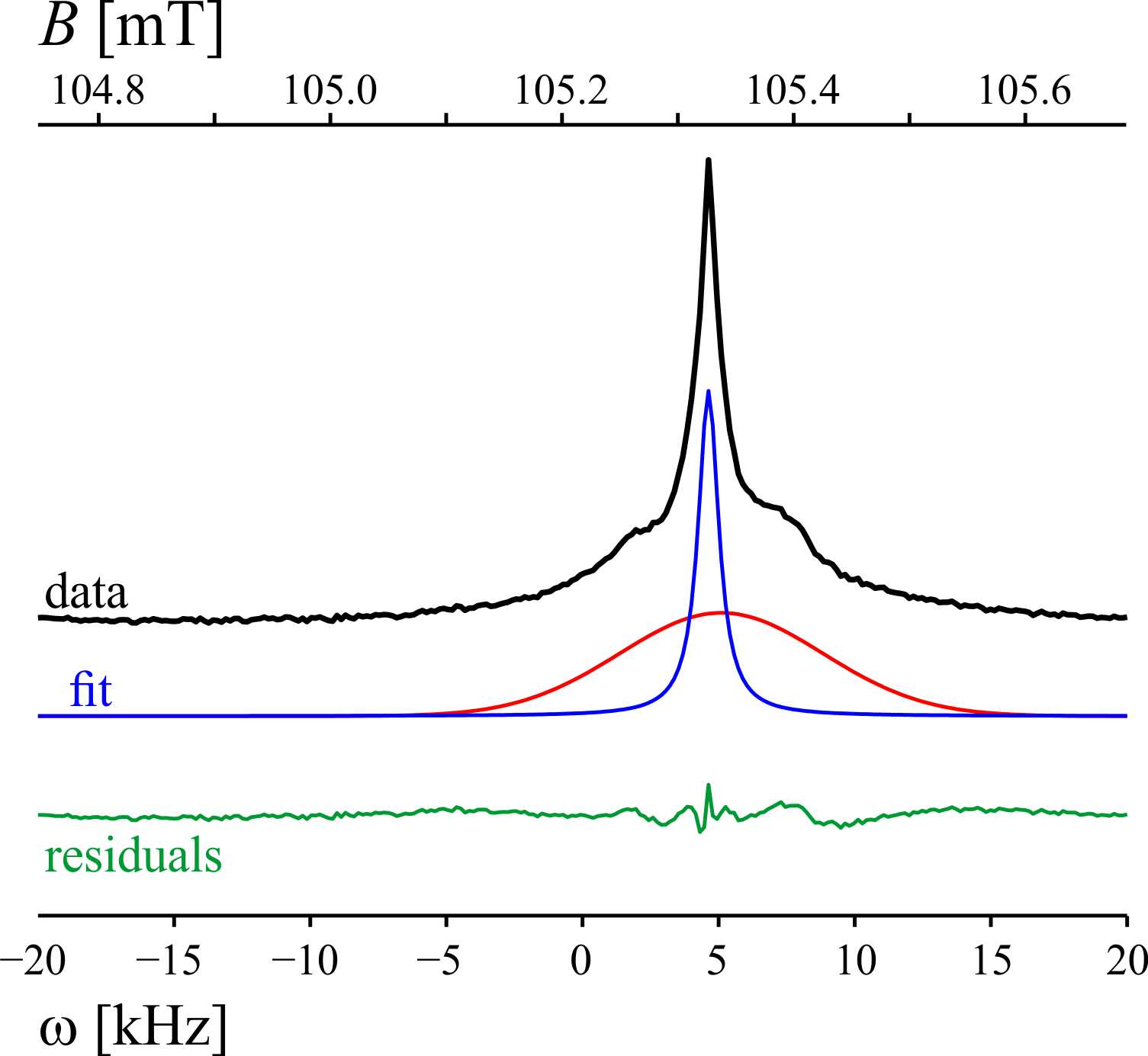}
  \caption{Black line: $^{1}$H-NMR-spectrum of a sample of pure water in the center of the magnet. (volume $10\times10\times5\,\text {mm}^3$, excitation frequency = 4.48 MHz, repetition time = 3 s, sampling: 256 points of 50\,$\mu$s duration, 64 co-added scans. The spectrum can be decomposed in a Lorentzian (blue line, with half height full width (HHFW) = 906\,Hz or 202\,ppm, centered at $\omega$ = 4.48463\,MHz or $B$ = 105.33\,mT) and a Gaussian (red line, with HHFW = 8765\,Hz or 1957\,ppm centered at $\omega$ = 4.48510 MHz or $B$ = 105.34 mT). The ratio integrals are Gauss/Lorentz = 2.1. The green line are the residuals from the fit (the ordinate of all figures has the same scale but different offsets)}
    \label{fig:6}
\end{figure}

To improve the characterization, we performed $^{1}$H-NMR experiments on a 5\,mL water sample contained in a 3D printed squared well with a $10\times10\,\text{mm}^2$ footprint. A solenoidal coil was used to apply the excitation field (axis along $z$, ID = 30\,mm, length 20\,mm, 17 windings of 1\,mm varnished copper wire). The coil was tuned and matched to a frequency of 4.48\,MHz. The result is shown in Fig. \ref{fig:6} (black line) and can be decomposed into a Lorentzian corresponding to $B=\left(105.33\pm 0.02\right)$\,mT and a Gaussian component corresponding to $B=\left(105.34\pm 0.21 \right)$\,mT. The width of the Lorentzian is a measure for the homogeneity. The fit parameters are given in the caption of Fig.\,\ref{fig:6} and result over the sample volume in an upper limit of 202\,ppm homogeneity.

\subsection{Using a diamond sensor}\label{ddiamond}
The purpose of this device is to provide the environment for a novel microwave-free diamond based magnetic field camera capable of investigating areas of interest up to cm$^2$ as illustrated in Fig. \ref{fig:GSLACSchematic}. Microwave-free diamond sensors considered here require a magnetic background field of 102.4\,mT \cite{wickenbrock2016microwave,levelanticrossing,Zheng2020}. We characterize the homogeneity of the constructed Halbach array additionally with a diamond-based sensor which offer higher spatial resolution than the other two types of sensors above. For this purpose we constructed a small fiberized NV-ensemble sensor, mounted it on a 3D-translation stage and measured the magnetic field in a $30\times20\,\text{mm}^2$ plane nearly perpendicular to the main magnetic field direction in the center of the Halbach array in steps of 1\,mm and 1.5\,mm in $z$ and $y$-direction, respectively. The focus of this manuscript is the construction of the magnet; here, we refrain from too many details about the NV center and its magnetometry methods. This is covered extensively in the literature, for example \cite{Budker2008NVcenter,PhysRevApplied.10.034044,DOHERTY2013NVCenter,Rondin_2014NVMag,2019BarryNVReview} and references therein. Additionally, another manuscript about the details of this sensor and the vector magnetometry method used here is in preparation. In the following, before presenting a highly-resolved magnetic field map of the magnet as the prime result of this section, we briefly summarize the main characteristics of the diamond sensor and describe the experimental setup and measurement procedure. This involves the reconstruction of the 3D magnetic field from optically detected magnetic resonance (ODMR) measurements. The results confirm the homogeneity of the magnetic field in the Halbach to be consistent with the Hall probe and the NMR measurement, but with a threefold improved field resolution and the additional vector information of the magnetic field.

\subsubsection{Diamond sensor setup}
\begin{figure*}
    \centering
    \includegraphics[width=1\textwidth]{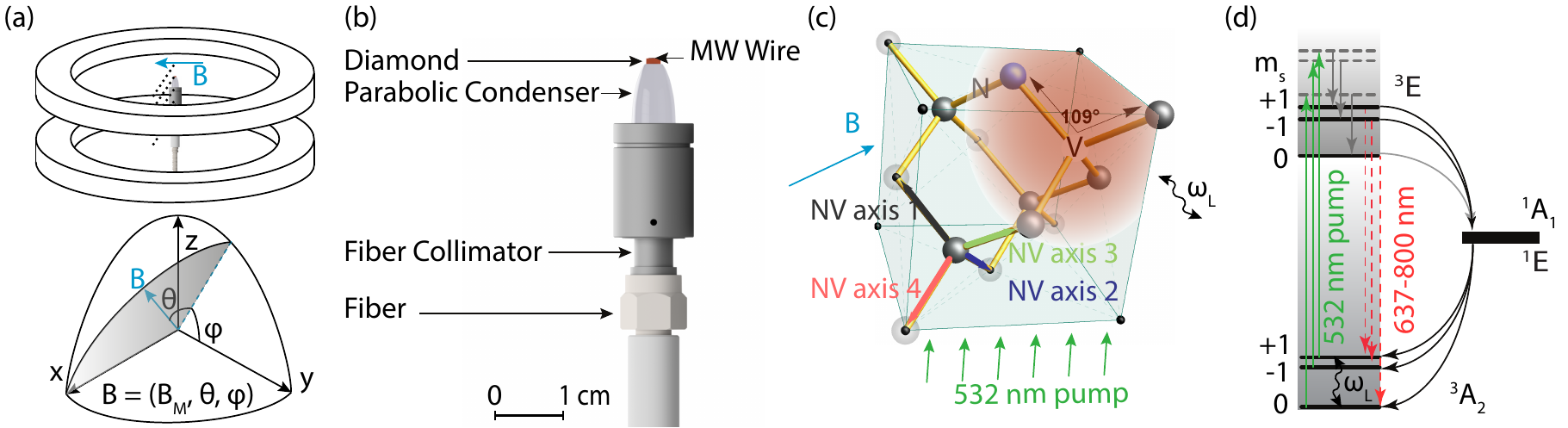}
    \caption{Diamond sensor schematic. a) The orientation of the diamond sensor in the Halbach array. The orientation of the magnetic field is given in spherical coordinates with angle $\theta$ being the latitude angle with respect to the x-axis and $\phi$ the longitude. The main component of the magnetic field points along the x-axis. b) Schematic of the fiberized sensor (see main text for details). c)\,Illustration of the diamond lattice structure indicating the four different NV axes 1,2,3,4 and their relation to the applied magnetic field B. Microwaves (MW) with frequency $\omega_L/\left(2\pi\right)$ are applied to the sample. d) Relevant energy level of the NV center.}
    \label{fig:diamondsetup}
\end{figure*}
The orientation of the fiberized sensor in the magnet is shown in Fig. \ref{fig:diamondsetup}a. The magnetic field is given in spherical coordinates $\vec{B}=\left(B_M,\theta,\phi\right)$. The angle between the magnetic field vector and the zy-plane is $\theta$, and $\phi$ is the angle between the projection of $\vec{B}$ in the zy-plane and the y-axis.
The diamond sample used for this sensor is a type Ib, 99.99\% $^{12}C$, (100)-cut, high-pressure, high temperature (HPHT) grown sample (from Element Six). The initial nitrogen concentration of the sample was specified as $\sim$\,13\,ppm, with an NV concentration of $\sim$\,3.7\,ppm. The diamond dimensions are $2\times2\times0.5\,\text{mm}^3$.
A schematic of the fiberized magnetic field sensor is shown in Fig\,\ref{fig:diamondsetup}b. The diamond is mounted on a  parabolic condenser lens, the concentrator itself is glued onto a fiber collimator (Thorlabs, F220SMA-532), which is screwed onto a high-power multi-mode fiber (Thorlabs, MHP365L02). On top of the diamond, a microwave (MW) wire loop is attached to provide oscillating magnetic fields required for this magnetic-field detection scheme. The green light illuminating the NV centers is provided by a diode-pumped solid-state laser (Laser Quantum, gem 532). It polarizes the NV center spin into the $m_s=0$ state within the triplet ground-state manifold. Figure \ref{fig:diamondsetup}d shows the relevant energy levels. The light is delivered to the diamond via compound parabolic concentrator, collimator and the high power fiber. The same components collect the spin-state dependent red photoluminescence (PL) of the NV ensemble. The PL is filtered by a longpass dichroic filter (Thorlabs, DMLP605) at the other end of the fiber and focused onto a photodiode (Thorlabs, PDA36a2) which is connected to a data acquisiton system and a lock-in amplifier (Stanford Research, SR830).
The MWs to manipulate the NV spins and drive transitions within the ground state energy levels are generated by a MW generator (SRS, SG396), amplified with a 16\,W amplifier (RF Lambda) and passed through a circulator (Minicircuits, CS-3.000), before they are applied to the NV centers.
To filter low-frequency systematic noise components, for example by laser power fluctuations, the MW-frequency $\omega_L$ is modulated and the detected PL demodulated with the lock-in amplifier. 
The sensor assembly is mounted on a computer-controlled motorized 3D translation stage (Thorlabs, MTS25/M-Z8) in the center of the two Halbach rings.

\subsubsection{Measurement procedure and results}
\begin{figure}
    \centering
    \includegraphics[width=1\columnwidth]{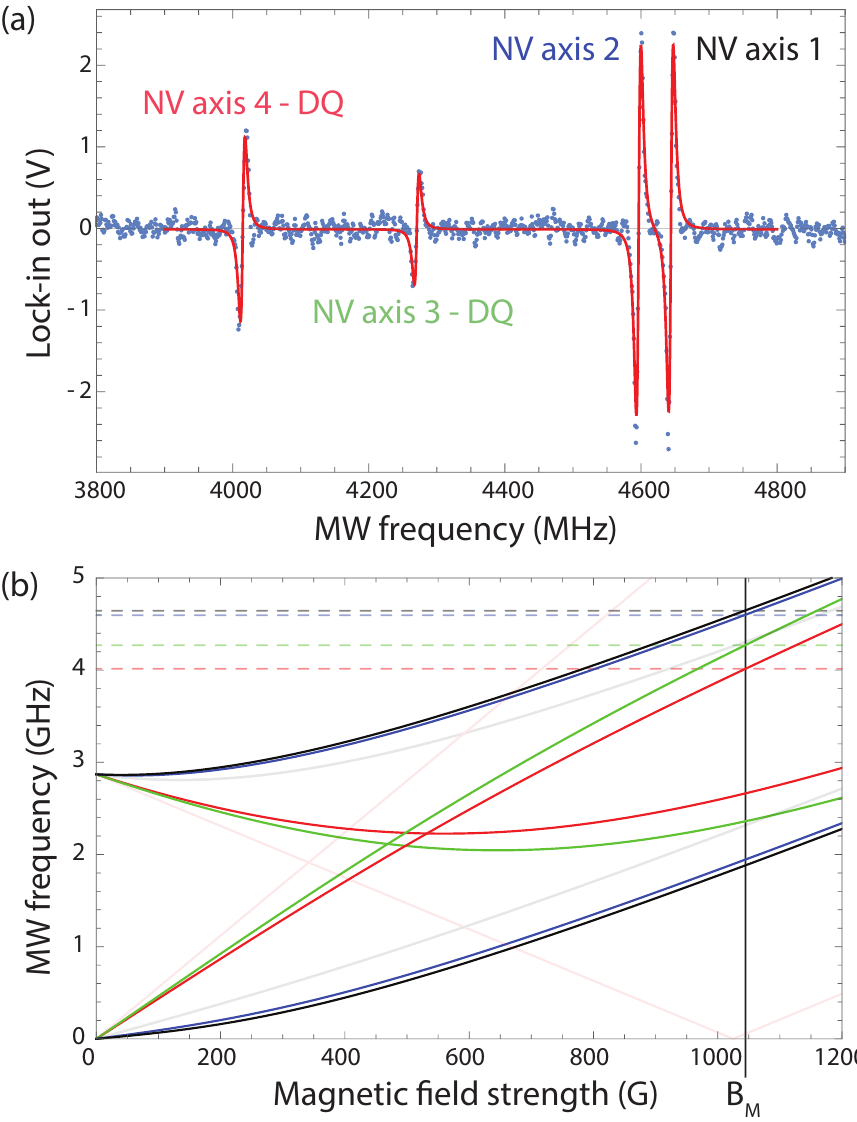}
    \caption{(a) Example data trace (blue) of four optically detected resonances at a given position near the center of the Halbach magnet. The resonances appear dispersive due to the use of a lock-in amplifier. The resonance are fitted (red) and their center frequencies are extracted. (b) Transition frequencies of all four axis in the diamond (axis 1,2,3,4 in black, blue, green, red, respectively) for a specific magnetic field direction as a function of the magnitude of the applied magnetic field. Translucent in the background are the transition frequencies when the magnetic field is applied along one of the diamond axis. The set of curves starting from the origin corresponds to double quantum transition for the respective NV axis. The center frequency fit result from (a) is indicated by horizontal lines in the appropriate color. This way the absolute magnetic field B$_\text{M}$ was extracted.}
    \label{fig:fieldFreq}
\end{figure}

At a given position the MW frequency is scanned (between 3800\,MHz and 5000\,MHz and the lock-in signal recorded. An example of the collected data is shown in Fig. \ref{fig:fieldFreq}a.

\begin{figure*}
    \centering
    \includegraphics[width=\textwidth]{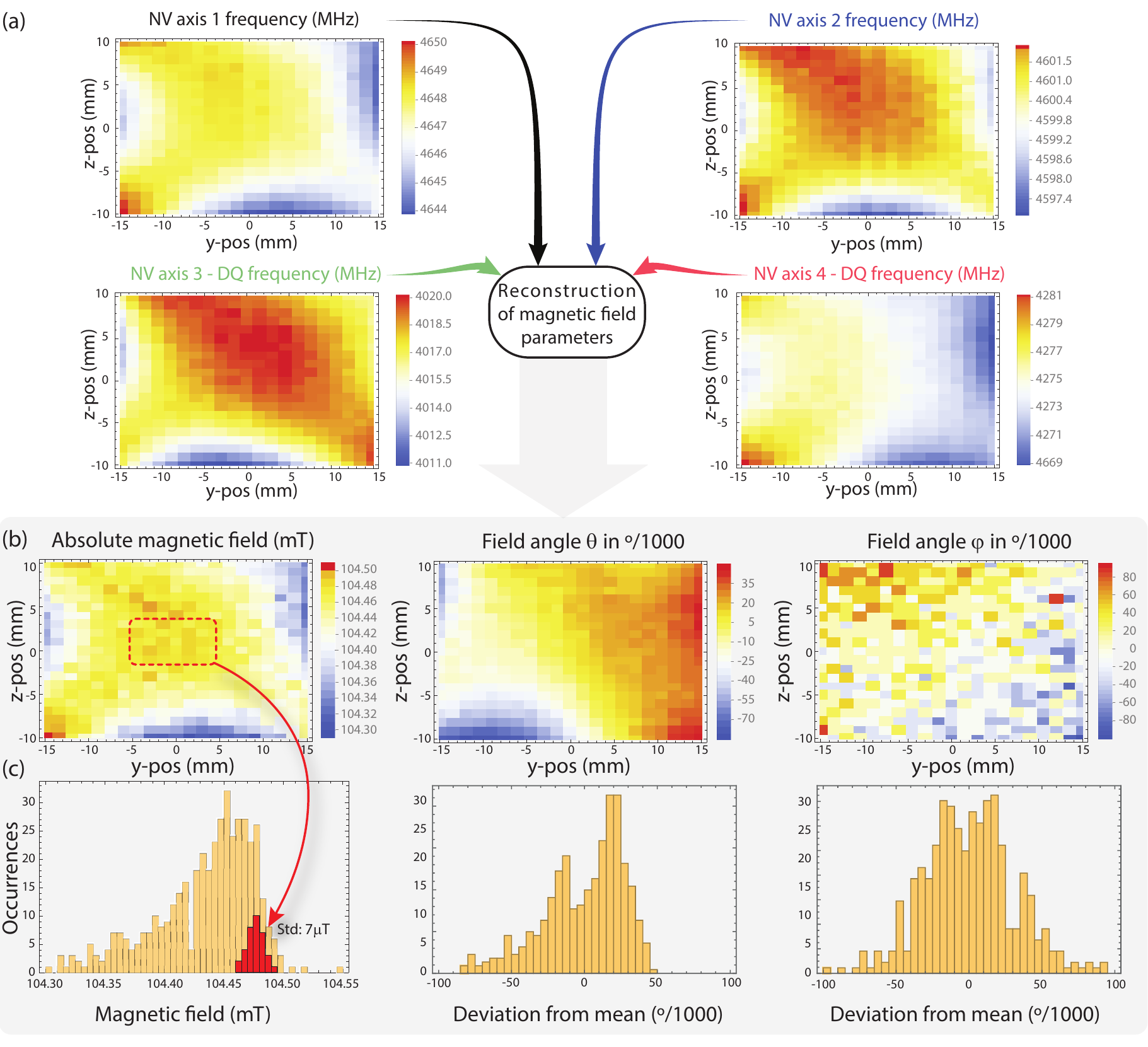}
    \caption[Frequency and field maps]{(a) Spatially resolved frequency maps of the individual resonances from Fig. \ref{fig:fieldFreq}a. (b) The data collected are used to reconstruct the vector magnetic field in spherical coordinates as defined in Fig.\,\ref{fig:diamondsetup}a. (c) Histograms of the data in (b). The data in red corresponds the absolute magnetic field values in the central 10\,$\times$\,5\,mm$^2$ of the magnet. Its standard deviation is 7\,$\mu$T.}
    \label{fig:results}
\end{figure*}

Four features can be identified in the demodulated lock-in data and are fitted with dispersive Lorentzian lineshapes with the center frequency as a fit parameter. All fit parameters are recorded with the position of the translation stage before the sensor is moved and the procedure repeated. Overall the sensor was moved in steps of 1\,mm and 1.5\,mm in $z$ and $y$-direction, respectively, such that an area of $20\times\,30\,\text{mm}^2$ was scanned. The resulting frequency maps can be seen in Fig.\,\ref{fig:results}, and they show structures corresponding to the strength of the longitudinal and transverse component of the magnetic field for a given diamond lattice axis.
The vector magnetic field at the position of the sensor can be reconstructed using this measurement. A manuscript detailing the procedure is currently in preparation. Here we want to sketch the mechanism: The relevant energy level scheme can be seen in Fig. \ref{fig:diamondsetup}d. Illuminating the diamond with green light initializes the NV centers along all four axis into the $m_s=0$ sublevel of the triplet ground state \cite{Doherty2013NV}. The magnetically sensitive sublevel $m_s=\pm1$ are offset by an energy corresponding to approximately 2870\,MHz, if there is no field applied. If microwaves are added with this frequency, population is transferred to the magnetically sensitive sublevels which exhibit a lower red photoluminescence rate under green illumination. As mentioned before, vector magnetic fields can also be sensed without using microwaves \cite{vectormagnetometer2}, however, a precise background field corresponding to the groundstate level anticrossing would be required. In its current state, the magnetic field of the  homogeneity-optimized Halbach magnet is around 2\,mT higher than the GSLAC field and therefore conventional vector magnetometry with microwaves was performed.

The observable features seen in the microwave spectroscopy in Fig.\,\ref{fig:fieldFreq} correspond to transitions from $m_s=0\rightarrow m_s=\pm1$, if the magnetic field is along the NV axis (longitudinal); if there is an additional magnetic field component orthogonal to the NV axis (transverse), transitions between the $m_s=-1\rightarrow m_s=+1$ are allowed as well. We call these double quantum (DQ) transitions.
For a given magnetic field direction at the position of the sensor, both, the longitudinal and transverse components of the magnetic field for each NV axis and therefore the energy level structure of this particular NV direction can be calculated. We parameterized the magnetic field in spherical coordinates (Fig.\,\ref{fig:diamondsetup}a), with an absolute magnetic field $B_M$ and two angles $\theta$ and $\phi$. While the absolute magnetic field $B_M$ is the same for all NV axes in the diamond, both angles differ depending on the specific axis (illustrated in Fig.\,\ref{fig:diamondsetup}c). Figure\,\ref{fig:fieldFreq} shows an example of all allowed microwave frequencies for all axis in a diamond as a function of the absolute magnetic field. The dashed lines correspond to measured microwave frequencies. With these the absolute field B$_M$ and the field angles $\theta$ and $\phi$ of the field with respect to the diamond lattice orientation can be extracted. 
Figure \ref{fig:results} contains a summary of the measurement result. The sensor was moved in steps of 1\,mm 1.5\,mm in a plane ($y$-$z$ plane) (nearly) perpendicular to the main magnetic field direction ($x$-axis) around the center of the two Halbach rings. The $30\times20\,\text{mm}^2$-maps of frequency measurements are shown in Figure \ref{fig:results}a. The four frequencies at each position are then used to numerically derive values of the absolute magnetic field B$_M$ and the two angles characterizing the field orientation with respect to the diamond. The resulting maps can be seen in Fig. \ref{fig:results}b. The average of angles $\theta$ and $\phi$ are $-2.433^\circ$ and $35.764^\circ$, respectively, which is consistent with the orientation of the diamond lattice (cf. Fig. \ref{fig:diamondsetup}c) relative to the coordinate system of the magnet. The angle maps show the deviation with respect to the mean in millidegree. Sub-figures in Fig.\,\ref{fig:results}c show histograms of all the extracted values of each map. In particular, the first histogram shows the magnetic field values corresponding to that in Fig.\,\ref{fig:results}b. The data from the central $10\times5\,\text{mm}^2$ region are highlighted in red. The standard deviation of these data corresponds to 7\,$\mu$T (or 67\,ppm) which is consistent with the measurement of the NMR linewidth. The average of the field value in the central region is 104.4770(7)\,mT, which is slightly lower than the value measured with NMR. We attribute this difference to temperature effects. Noteworthy is also the measurement accuracy of the relative angle between field and diamond lattice. The standard deviation of angles $\theta$ and $\phi$ of the central region data (35 measurements) is $9\times10^{-3}\,^\circ$ and $18\times10^{-3}\,^\circ$, respectively. 
The overall pattern of the reconstructed magnetic field in in good agreement with numerical simulations of the magnet.

\section{Conclusion and Outlook}
We designed, built and characterized a novel permanent magnet-based Halbach array providing a highly homogeneous magnetic field of predefined strength in its center and over a large volume. This magnet is going to be used in a microwave-free magnetic imaging setup which informs the parameter requirements on homogeneity and background field strength.  
The novel design for the double Halbach ring based on cylindrical magnets has several important advantages over conventional Halbach magnets built before. The use of cylindrical magnets provides better homogeneity per size or mass of the magnet. The suggested concept to orient the cylindrical PMs by orientation magnets is an additional step in the already quite elaborated construction of Halbach magnets. However, the homogeneity of the final magnet turned out to be very satisfactory. Considering the strong temperature dependence of FeNdB-permanent magnets, the simulated field was met very accurately (104.5\,mT). In order to tune the field exactly to 102.4\,mT as required by the microwave-free sensing scheme, a resistive coil system will be implemented. Additionally, the final homogeneity of 67\,ppm is good agreement to the simulations (9\,ppm) without much fine tuning or shimming. From experience two orders of magnitude discrepancy between experiment and simulation can be expected. We attribute this improvement to a) the narrow $B_R$-distribution of the PMs, b) the precise orientation of the PMs before assembly, and c) encasing the PMs in a protective construction made from aluminium. The latter avoids mechanical impact of magnets, which could partially demagnetize the PMs. We like to note that the design is scalable and much stronger magnets can be built (e.g. for classical NMR).

The magnet fulfills all the requirement to perform microwave-free wide-field imaging experiments using large diamonds with an implanted NV layer. The small fiberized diamond sensor used to characterize the magnet showed impressive characteristics in terms of spatial resolution (mm), field resolution (7\,uT) and angular resolution of the magnetic field vector ($9\times10^{-3}\,^\circ$ and $18\times10^{-3}\,^\circ$, for angles $\theta$ and $\phi$, respectively). We note here, that the displacement due to a rotation of the actual diamond sensor with an assumed volume of $500^3\,\mu\text{m}^3$ by an angle corresponding to the angular resolution would be well below the diffraction limit of conventional microscopy. The small microwave-requiring diamond sensor will be replaced with a larger diamond sample to perform microwave-free magnetic field imaging.

 The expected field-of-view is potentially much larger than achievable with conventional (microwave-requiring) ODMR measurements due to the high homogeneity of the background field. Additionally, without the need for microwaves, conductive materials and samples with a high dielectric loss (like water) can also be magnetically imaged. Potential applications of this non-invasive technique lie in biology, geology and materials science. First experiments of this microwave-free technique are in preparation.

\section{Acknowledgment}
This work is supported by the EU FET-OPEN Flagship Project
ASTERIQS (action 820394), the German Federal Ministry of Education and Research (BMBF) within the Quantumtechnologien program (grants FKZ 13N14439 and FKZ 13N15064), the Cluster of Excellence “Precision Physics, Fundamental Interactions, and Structure of Matter” (PRISMA+ EXC 2118/1) funded by the German Research Foundation (DFG) within the German Excellence Strategy (Project ID 39083149).
We thank Dmitry Budker and Muhib Omar for informative discussions, helpful advice and generous support.

\bibliographystyle{ieeetr}
\bibliography{Literature.bib}

\end{document}